\documentclass[amssymb,aps,twocolumn,showpacs,nofootinbib]{revtex4}
\usepackage[]{graphicx}
\def\ket#1{| #1 \rangle}
\def\bra#1{\langle #1 |}
\def\braket#1#2{\langle #1 | #2 \rangle}

\begin{document}

\title{Unconditionally secure key distillation from multi-photons}

\author{Kiyoshi Tamaki}
\author{Hoi-Kwong Lo}
\affiliation{Center for Quantum Information and Quantum Control, Dept. of Electrical \& Computer Engineering and Dept. of Physics, University of Toronto, Toronto, Ontario, M5S 3G4, CANADA}


\begin{abstract}
In this paper, we prove that the unconditionally secure key can be surprisingly extracted from {\it multi}-photon emission part in the photon polarization-based QKD. One example is shown by explicitly proving that one can indeed generate an unconditionally secure key from Alice's two-photon emission part in ``Quantum cryptography protocols robust against photon number splitting attacks for weak laser pulses implementations'' proposed by V. Scarani {\it et al.,} in Phys. Rev. Lett. {\bf 92}, 057901 (2004), which is called SARG04. This protocol uses the same four states as in BB84 and differs only in the classical post-processing protocol. It is, thus, interesting to see how the classical post-processing of quantum key distribution might qualitatively change its security. We also show that one can generate an unconditionally secure key from the single to the four-photon part in a generalized SARG04 that uses six states. Finally, we also compare the bit error rate threshold of these protocols with the one in BB84 and the original six-state protocol assuming a depolarizing channel.
\end{abstract}


\maketitle

Quantum key distribution (QKD) allows two separate parties, the sender Alice and the receiver Bob, to share a secret key with negligible leakage of its information to an eavesdropper Eve. The best known QKD protocol is BB84 protocol published by Bennett and Brassard in 1984 \cite{BB84}. Many aspects of the BB84 protocol including the unconditional security \cite{M96,SP00,GLLP02} and its implementations \cite{GRTZ02} has been investigated. BB84 is unconditionally secure if Alice emits a single-photon. However, if Alice emits multi-photon, Eve in principle gets full information on bit values without inducing any bit error by exploiting a photon number splitting attack (PNS) \cite{BLMS00}. 

Recently, Scarani, et.al. \cite{SARG04} have proposed a QKD (SARG04) that is robust against PNS attack. This protocol uses exactly the same four states as the one in BB84, and only the classical data processing is different from BB84. A key goal of this paper is to demonstrate that among many modifications of BB84 \cite{LCA00}, SARG04 is the first essential modification in the sense that it has a property that BB84-type QKD has never accomplished, i.e., one may generate a secure key not only from the single-photon part, but rather surprisingly also from a two-photon part. In SARG04, the classical part is modified in such a way that after Alice's initial broadcast, the two remaining states are nonorthogonal. Thus, even by using PNS attack, Eve cannot discriminate the state deterministically. This is an intuition that one might expect to generate a secure key from the two-photon part. 

We remark that this kind of secure key distillation is natural from the viewpoint of an unambiguous state discrimination \cite{C01}. It is known that an unambiguous discrimination among $N$ states of a qubit space is only possible when at least $N-1$ copies of the state are available. This means in the case of four states that we have no chance to distill a key from more than the two-photon part, because if Eve succeeds the discrimination, then she can resend the corresponding state, while if she does not, then she sends vacuum state to Bob, which disguises for channel losses. In other words, there is no reason that forbids the generation of a secure key from {\it both} the single-photon and two-photon parts in a four-state protocol. By modifying only the classical part in BB84, SARG04 might accomplish this.

Note that SARG04 differs from BB84 {\it only} in the classical communication. Thus, it is very interesting to see how {\it only} the classical communication of QKD changes its security, which is a fundamentally interesting question. This is related to the viewpoint of `` Entanglement as precondition for secure QKD'' in \cite{CLL03}. So far, many studies have been done to generate a single-photon source in experiments for QKD \cite{GRTZ02}. Hence, the demonstration of a secure key from the two-photon part has an impact to that direction of studies. Moreover, from practical viewpoint, an experiment for SARG04 should not be so difficult once an experiment for BB84 is available (e.g. see \cite{C04}). It follows that to investigate which protocol one should perform is important from the practical viewpoint. In summary, to prove the unconditional security of SARG04 is an interesting question both from fundamental and practical viewpoints.

In this paper, we prove that the unconditionally secure key can be surprisingly extracted not only from single-photon part, but also from {\it multi}-photon part in the photon polarization-based QKD, especially two-photon part in the SARG04 protocol. Thus, our result clearly shows that the modification of {\it only} the classical communication part in QKD can change its quality. In this paper, we assume that Alice has a coherent light source and Bob has a single-photon detector with no dark count. To prove the security of the two-photon part, we generalize the idea of ``squash operation'' in \cite{GLLP02}, where the authors treat the multi-photon part just by assuming the worst case scenario for BB84. In our case, we cannot rely on this scenario, because this scenario completely denies a chance to generate a secure key form the two-photon part. The generalization we made can widely apply to the multi-photon parts in most of polarization based QKD, such as a modified SARG04 protocol (based on six states) that we propose in this paper. In this protocol a secure key can be generated from the single to the four-photon part.

This paper consists of the following. We first present our notations and describe how SARG04 works. After that, we prove the security of the protocol with the single-photon part and two-photon part. Finally, we compare the security of SARG04 with the one of BB84, and we end this paper with mentioning a natural generalization of SARG04 followed by a summary.

We first define several notations. $\{\ket{0_x},\ket{1_x}\}$ is a $X$-basis for a qubit, which is related to $Z$-basis and $Y$-basis by $\{\ket{j_z}\equiv[\ket{0_x}+(-1)^j\ket{1_x}]/\sqrt{2}\}$ ($j=0,1$) and $\{\ket{j_y}\equiv[\ket{0_x}+i(-1)^j\ket{1_x}]/\sqrt{2}\}$, respectively. We define a filtering operator $F=\sin\frac{\pi}{8}\ket{0_x}\bra{0_x}+\cos\frac{\pi}{8}\ket{1_x}\bra{1_x}$, a $\pi/2$ rotation around $Y$-axis $R\equiv\cos\frac{\pi}{4}\openone_{\rm qubit}+\sin\frac{\pi}{4}(\ket{1_x}\bra{0_x}-\ket{0_x}\bra{1_x})$, and $\ket{\varphi_{j}}\equiv\cos\frac{\pi}{8}\ket{0_x}+(-1)^j\sin\frac{\pi}{8}\ket{1_x}$, where $\openone_{\rm qubit}\equiv\sum_{j=0}^{1}\ket{j_x}\bra{j_x}$. Note that $R\ket{\varphi_{1}}=\ket{\varphi_{0}}$. We introduce $\hat P(X\ket{\Psi})\equiv X\ket{\Psi}\bra{\Psi}X^{\dagger}$, $\ket{\chi_{0\pm}}=\frac{1}{\sqrt2}(\ket{0_z}\ket{0_z}\pm\ket{1_z}\ket{1_z})$, $\ket{\chi_{1\pm}}=\frac{1}{\sqrt2}(\ket{0_z}\ket{1_z}\pm\ket{1_z}\ket{0_z})$, $\ket{\nu, \varphi}\equiv\ket{\varphi}^{\otimes\nu}$, and $\ket{\overline\varphi_{j}}$ satisfing $\braket{\overline\varphi_{j}}{\varphi_{j}}=0$.

\begin{figure}[tbp]
\begin{center}
 \includegraphics[scale=0.6]{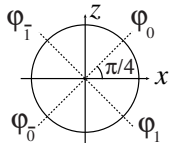}
\end{center}
 \caption{Bob's measurement basis in the Bloch sphere. Note that the random rotation $R^{K'}$ just changes the definition of the outcomes and does not change the bases as a set. \label{fig1}}
\end{figure}

We now explain how SARG04 works. Since this protocol is similar to B92 protocol \cite{B92}, we explain SARG04 in the context of the modification of B92 protocol. Imagine the B92 where Alice randomly sends $\ket{\nu, \varphi_j}$ ($j=0,1$) depending on the bit value $j$, while Bob performs the B92 measurement where he randomly chooses a measurement basis from $\{\ket{\varphi_{j'}},\ket{\overline{\varphi_{j'}}}\}$ ($j'=0,1$) (see also Fig.~\ref{fig1}). If his measurement outcome is $\overline{\varphi_1}$ or $\overline{\varphi_0}$, which we call conclusive, then Bob broadcasts that he got a conclusive results. From the outcome, he can infer which bit value Alice sent to him, i.e., when the outcome is $\overline{\varphi_1}$ ($\overline{\varphi_0}$), he concludes that Alice sent bit value $0$ ($1$). 

We can convert the above B92 into SARG04 just imposing Alice to perform a rotation $R^{K}$ just before sending the state, and imposing Bob to perform a rotation $R^{-K'}$ just before performing the B92 measurement. Here, each of $K$ and $K'$ is randomly chosen from $0$ to $3$ ($R^{0}\equiv \openone_{\rm qubit}$). After Bob performs the measurement, Alice broadcasts to Bob which B92 she has chosen, i.e., she broadcasts $K$. If $K=K'$, then Bob broadcasts whether the measurement outcome is conclusive or not, and if $K\neq K'$, then they discard all corresponding data. It is easy to see that Alice and Bob perform the same operations as in BB84 (see also Fig.~\ref{fig1}). Intuitively, the symmetrization of a quantum channel, including Eve's action, given by the random rotation $R$ provides an advantage to SARG04 over the B92. Actually, we will prove an error threshold of SARG04 that is independent of quantum channel losses, which is a big difference from the case for the B92 \cite{TL04}.

Before proving the unconditional security of SARG04 with $\nu$-photon, we describe how we treat the case that Bob's measurement outcome is both $\varphi_{j'}$ and $\overline{\varphi_{j'}}$. This happens because of multi-photon detections or dark counts. In such a case, we impose Bob to decide his measurement outcome randomly. Note that Bob can equivalently do this by locally preparing a random qubit state followed by the measurement on it. We pessimistically assume that Eve prepares the qubit state instead of Bob. Since we can assume that Eve sends a qubit state in the non-ambiguity case, without threatening any security we consider Eve who always sends a qubit state or vacuum state to Bob. This process can be regarded as a ``squash operation'' \cite{GLLP02}.

In order to prove the security of SARG04, we construct an unconditionally secure Entanglement Distillation Protocol with $\nu$-photon (EDP-$\nu$) that can be converted to SARG04 with $\nu$-photon. This protocol employs an EDP \cite{BDSW96} based on Calderbank-Shor-Steane (CSS) codes \cite{SP00,CSS}. In EDP-$\nu$, Alice creates many pairs of qubits in the state $\ket{\Psi^{(\nu)}}_{\rm AB}=\frac{1}{\sqrt2}(\ket{0_z}_{\rm{A}}\ket{\nu, \varphi_0}_{\rm{B}}+\ket{1_z}_{\rm{A}}\ket{\nu, \varphi_1}_{\rm{B}})$, and after randomly applying the rotation $R^{K}$ to the system B, Alice sends the system B to Bob. On the other hand, Bob randomly applies the rotation $R^{-K'}$ to the qubit state, and then he performs a filtering operation whose successful and failure operation is described by Kraus operators $F$ and $\sqrt{\openone_{\rm qubit}-F^2}$, respectively. After many repetition of state sending and Bob's operation, they use classical communication so that they keep the qubit pairs where Bob's filtering succeeds with $K=K'$. From these pairs, they randomly choose test pairs that are subjected to measurements in the $Z$-basis by Alice and Bob. Thanks to random sampling theorem, the test pairs give us a good estimation of the bit error rate on the remaining pairs (code pairs) provided that the number of the test and code pairs are large enough. If they can estimate the upper bound of the phase error rate on the code pairs, they can choose CSS codes that correct both bit and phase errors on the code pairs so that they share some maximally entangled states in the form of $\ket{\chi_{0+}}$. Finally, by performing $Z$-basis measurement on those states, they share a secret key. 

To confirm that EDP-$\nu$ is completely equivalent to SARG04, first note that Alice can perform $Z$-basis measurement just before sending the system B without changing any measurement outcome. It follows that Alice randomly sends $\ket{\nu, \varphi_j}$ ($j=0,1$), and this is exactly what Alice does in SARG04. Similarly, we can allow Bob to perform $Z$-basis measurement just after the filtering operation, which is completely the same as the measurement randomly chosen from $\{\ket{\varphi_{j'}},\ket{\overline{\varphi_{j'}}}\}$ basis ($j'=0,1$). This can be seen by noting that $F\ket{{j'}_z}\bra{{j'}_z}F^{\dagger}=\frac{1}{2}\ket{\overline{\varphi_{j'}}}\bra{\overline{\varphi_{j'}}}$. By combining the random rotation, it is obvious that Bob's operation in EDP-$\nu$ is completely the same as the one Bob does in SARG04. Note that successful filtering events corresponds to the conclusive events.

Since we have seen the equivalence of EDP-$\nu$ to SARG04, we prove the security of SARG04 based on EDP-$\nu$. Note that the bit error rate on the code pairs is well estimated by the test pairs, hence all we have to consider is how to estimate the phase error rate on the code qubit pairs from the bit error rate. Intuitively, this phase error estimation is given by the symmetry of the rotations $R$, and the property of the filtering operation $F$ \cite{BTBLR04,TL04}. Let us define $p_{L, \nu}^{(l)}$ ($L=\{{\rm Bit}, {\rm Phase}, {\rm Fil}\}$) as an expectation value for a particular $l^{th}$ qubit pair of the $\nu$-photon part having an event in $L$, conditioned on arbitrary configurations of an event in $L$ or the failure filtering including Bob's vacuum detection for the previous $l-1$ pairs. Furthermore, let us define a random variable $X_{L, \nu}^{s}\equiv n_{L, \nu}^{s}-\sum_{l=1}^{s}p_{L, \nu}^{(l)}$, where $n_{L, \nu}^{s}$ is the number of events $L$ with $\nu$-photon {\it actually} has happened from $1^{st}$ pair to $s^{th}$ pair. By directly applying Azuma's inequality \cite{A67} to $X_{L, \nu}^{s}$, one can show that $\sum_{l=1}^{s}p_{L, \nu}^{(l)}\rightarrow n_{L, \nu}^{s}$ with exponentially as the number of pairs $s$ increases. Thus, we have the following theorem.

{\it THEOREM: If $Cp_{\rm bit, \nu}^{(l)}+C'p_{\rm fil, \nu}^{(l)}\geqslant p_{\rm ph, \nu}^{(l)}$ holds, then $Ce_{\rm bit}^{(\nu)}+C'\geqslant e_{\rm ph}^{(\nu)}$ is exponentially reliable as the number of successfully filtering states increases. Here $e_{bit/ph}^{(\nu)}$ is the {\it actual} bit/phase error rate normalized by the actual successful filtering events.}

We emphasize that thanks to Azuma's inequality this theorem holds even when Eve performs the most general attack, including coherent attacks. Our theorem is a generalization of the discussion in \cite{BTBLR04}. Now, we are only left to obtain the inequality for a particular qubit pair in the form of $Cp_{\rm bit, \nu}+C'p_{\rm fil, \nu}\geqslant p_{\rm ph, \nu}$.

\begin{figure}[tbp]
\begin{center}
 \includegraphics[scale=0.5]{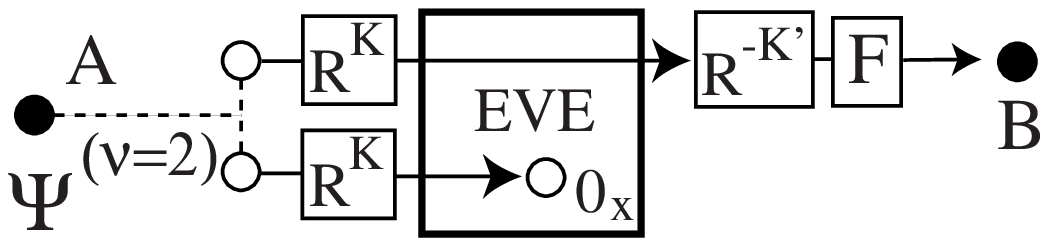}
\end{center}
 \caption{In the two-photon case (i.e., $\nu =2$), Alice first prepares three-qubit in the state of $\ket{\Psi^{(\nu=2)}}_{\rm AB}$. After some operations by Alice and Bob, they try to distill a key from a final state of the system ${\rm A}$ and ${\rm B}$ (black dots) if $K=K'$. 
 \label{fig2}}
\end{figure}

We remark that the pessimistic assumption on the state Eve sends to Bob and the above theorem are important observations. With these observations, we are left only to calculate the state of a qubit pair state, and find the relationship that holds for any Eve's action, which are straight forward. Moreover, to simplify Eve's action, we define ``trash'' systems that are qubits originated from multi-photon, but Bob has no interest in. Since Bob never care about the state of ``trash'' after Eve's action, we can safely assume that the final state of each trash is in a particular state, say $\ket{0_x}_{\rm trash}$ (see also Fig.~\ref{fig2}). Since we have put no assumption on Alice and Bob other than they use qubits, our basic strategy for the security proof can widely apply to any photon number part in most of polarization based QKD. 

For the later convenience, let $\rho_{\rm qubit}^{(\nu)}$ be the pair qubit state stemming from the $\nu$-photon part. With this state, $p_{L, \nu}$ is expressed as $p_{\rm fil, \nu}={\rm Tr}\Big[\rho_{\rm qubit}^{(\nu)}\Big]$, $p_{\rm bit, \nu}={\rm Tr}\Big[\rho_{\rm qubit}^{(\nu)}\sum_{m=+,-}\hat P(\ket{\chi_{1,m}})\Big]$, and $p_{\rm ph, \nu}={\rm Tr}\Big[\rho_{\rm qubit}^{(\nu)}\sum_{m'=0,1}\hat P(\ket{\chi_{m',-}})\Big]$. To obtain $\rho_{\rm qubit}^{(\nu)}$, we first consider the final state of $\ket{\Psi^{(\nu)}}_{\rm AB}$ after Alice, Bob, and Eve's operations with $K=K'$. This state is obtained by tracing out the every other pair from the total state, to which Eve has performed an arbitrary operation, including the one in coherent attacks. The final (unnormalized) state can be expressed as $\rho_{\rm fin}^{(\nu)}=\sum_{f}\rho_{\rm fin}^{(f,\nu)}$, where $f$ is an index for an arbitrary matrix representing Eve's action $E_{\rm B}^{(f,\nu)}$ on $\nu$-photon \cite{Note}.

In the single-photon case (i.e., $\nu =1$), $\rho_{\rm fin}^{(\nu)}=\rho_{\rm qubit}^{(\nu=1)}$ and $\rho_{\rm fin}^{(f,\nu=1)}=\rho_{\rm qubit}^{(f,\nu=1)}$. It is a bit tedious but straight forward to see that $p_{\rm ph, 1}=\frac{3}{2}p_{\rm bit, 1}$ for $\rho_{\rm qubit}^{(f,\nu=1)}$ stemming from any $E_{\rm B}^{(f,\nu=1)}$. Thus, by the linearity of the density matrix, we conclude that $e_{\rm ph}^{(1)}=\frac{3}{2}e_{\rm bit}^{(1)}$. Furthermore, one can also show that $\bra{\chi_{0-}}\rho_{\rm qubit}^{(\nu=1)}\ket{\chi_{0-}}\ge2\bra{\chi_{1+}}\rho_{\rm qubit}^{(\nu=1)}\ket{\chi_{1+}}$ and $2\bra{\chi_{1-}}\rho_{\rm qubit}^{(\nu=1)}\ket{\chi_{1-}}\ge\bra{\chi_{0-}}\rho_{\rm qubit}^{(\nu=1)}\ket{\chi_{0-}}$ always hold, which implies that there is a mutual information between phase and bit error patterns.

In the two-photon case ($\nu =2$), $\rho_{\rm qubit}^{(\nu=2)}$ and $\rho_{\rm qubit}^{(f,\nu=2)}$ are obtained by taking projection to $\rho_{\rm fin}^{(\nu=2)}$ and $\rho_{\rm fin}^{(f,\nu=2)}$ by $\ket{0_x}_{\rm trash}$ that can be expressed via a $2\times2$ matrix, $E_{\rm B}^{(f,u)}$ ($u=0,1$) \cite{Note2}. It is tedious but straight forward to see that if $y\geqslant g(x)\equiv\frac{1}{6}\left(3-2 x+\sqrt{6-6\sqrt{2}x+4x^2}\right)$ is satisfied, then $xp_{\rm bit,\nu=2}+yp_{\rm fil,\nu=2}\geqslant p_{\rm ph,\nu=2}$ holds for any $E_{\rm B}^{(f,u)}$ \cite{Note3}. Thus, we pessimistically conclude that $xe_{\rm bit}^{(2)}+g(x)=e_{\rm ph}^{(2)}$. Note that $e_{\rm ph}^{(2)}\neq0$ even when $e_{\rm bit}^{(2)}=0$ because ${\rm Inf}[g(x)]=\sin^2\frac{\pi}{8}$, which means Eve can get some information on the key without introducing any bit error in $\nu=2$ case. In $\nu=2$ case, we cannot find any mutual information between the bit and phase errors.

Since we have finished the phase error estimation, we can calculate the key generation rate for SARG04. By assuming the random hashing CSS code \cite{BDSW96, L01}, the key generation rate for $\nu$-photon part $R_{\nu}$ is asymptotically represented by $R_{\nu}=1-H(X_{\nu},Z_{\nu})$ \cite{L01}, where $H(X_{\nu},Z_{\nu})$ is the entropy of bit and phase error pattern in the $\nu$-photon part. By solving $R_{\nu}\geqslant0$, we can show that up to $e_{\rm bit}^{(1)}\sim 9.68\%$ (this is the same as the one recently obtained in \cite{BGKS05} without the ``preprocessing'') and $e_{\rm bit}^{(2)}\sim 2.71\%$ (when $x\sim2.747$) we can distill a secure key from the $\nu$-photon part. To compare the bit error rate threshold of SARG04 to the one of BB84, we assume that Eve simulates a depolarizing channel where a $\nu$-photon state $\rho^{\otimes\nu}$ evolves to $\left(1-\frac{4p}{3}\right)\rho^{\otimes\nu}+\frac{4p}{3}(\openone_{\rm qubit}/2)^{\otimes\nu}$. Here, $p$ is a depolarizing rate. Since $e_{\rm bit}^{(\nu)}=4p/(3+4p)$ in this channel, the single-photon and two-photon part of SARG04 is secure up to $p\sim8.04\%$ and $p\sim2.08\%$, respectively while BB84 with one-way classical communication is secure up to $p\sim16.5\%$ \cite{SP00}.

We can express a total key rate $R$ by using the decoy state \cite{LMC04}, which allows us to use an imperfect light source and imperfect threshold detectors. This idea gives the lower bound of the fraction of Bob's conclusive results conditioned on $\nu$-photon emission as $\xi(\nu)$ and the upper bound of the bit error as $\overline{e_{\rm bit}^{(\nu)}}$.  From them, we can compute the upper bound of the {\it conditional} entropy of the phase error pattern given the bit error patter, which is denoted by $\overline{H(Z_{\nu}|X_{\nu})}$. Hence, $R=-P_{\rm conc}h(e_{\rm bit})+\sum_{\nu=1}^{2}\xi(\nu)\Big(1-\overline{H(Z_{\nu}|X_{\nu})}\Big)$ \cite{GLLP02, L01, LMC04}. Here, $P_{\rm conc}$ is a fraction that Bob obtains the conclusive results and $e_{\rm bit}$ is the bit error rate on every conclusive result.

Note that our security analysis can directly apply to a modified six-state protocol, where Alice and Bob additionally perform a random $\pi/2$ rotation around $\{\ket{\varphi_{j'}},\ket{\overline{\varphi_{j'}}}\}$ axis in SARG04. By following the discussion on the unambiguous state discrimination, we expect that we may distill a secure key from the single to the four-photon part. Actually, one can show that we can indeed generate a secure key from $\nu$-photon part up to the error rates of $11.2\%$ ($\nu=1$), $5.60\%$ ($\nu=2$), $2.37\%$ ($\nu=3$), and $0.788\%$ ($\nu=4$), which correspond to $p\sim9.49\%$, $p\sim4.45\%$, $p\sim1.82\%$, and $p\sim0.595\%$, respectively, while $p\sim19.0\%$ in the original six-state protocol with one-way classical communication \cite{B98,L01}.

In this paper, we prove that the unconditionally secure key can be extracted from {\it multi}-photon emission part in the photon polarization-based QKD. Our result demonstrates clearly that by changing {\it only} the classical post processing protocol, the foundations of the security can change qualitatively.

We thank J.-C. Boileau, J. Batuwantudawe, M. Koashi and F. Fung for helpful discussions.

\end{document}